\documentstyle[12pt,aasms4]{article}

\def\arcsecpoint{$''\!.$}
\def\deg{$^{\rm o}$}

\received{}
\accepted{}

\lefthead{Crenshaw \& Kraemer}
\righthead{Extended Continuum in NGC 1068}

\begin{document}

\title{Resolved Spectroscopy of the Narrow-Line Region in NGC 1068. I.
The Nature of the Continuum Emission\altaffilmark{1}}

\author{D. Michael Crenshaw\altaffilmark{2}
\& Steven B. Kraemer\altaffilmark{3}}
\affil{Catholic University of America and Laboratory for Astronomy and
Solar Physics, NASA's Goddard Space Flight Center, Code 681,
Greenbelt, MD  20771}

\altaffiltext{1}{Based on observations with the NASA/ESA {\it Hubble Space 
Telescope}, which is operated by the Association of Universities for Research 
in Astronomy, Inc., under NASA contract NAS5-26555.}
\altaffiltext{2}{crenshaw@buckeye.gsfc.nasa.gov}
\altaffiltext{3}{stiskraemer@yancey.gsfc.nasa.gov}

\begin{abstract}
We present the first long-slit spectra of the Seyfert 2 galaxy NGC 1068 obtained 
by the Space Telescope Imaging Spectrograph (STIS); the spectra cover the 
wavelength range 1150 -- 10,270 \AA\ at a spatial resolution of 0\arcsecpoint05 
-- 0\arcsecpoint1 and a spectral resolving power of $\lambda$/$\Delta\lambda$ 
$\approx$ 1000. In this first paper, we concentrate on the far-UV to near-IR 
continuum emission from the continuum ``hot spot'' and surrounding regions  
extending out to $\pm$6\arcsecpoint0 ($\pm$432 pc) at a position angle of 
202\deg.

In addition to the broad emission lines detected by spectropolarimetry, the hot 
spot shows the ``little blue bump'' in the 2000 -- 4000 \AA\ range, which is 
due to Fe~II and Balmer continuum emission. The continuum shape of the hot spot 
is indistinguishable from that of NGC~4151 and other Seyfert 1 galaxies. Thus, 
the hot spot is reflected emission from the hidden nucleus, due to electron 
scattering (as opposed to wavelength-dependent dust scattering). The hot spot is 
$\sim$0\arcsecpoint3 in extent and accounts for 20\% of the scattered light in 
the inner 500~pc.

We are able to deconvolve the extended continuum emission in this region into 
two components: electron-scattered light from the hidden nucleus (which 
dominates in the UV) and stellar light (which dominates in the optical and 
near-IR).  The scattered light is heavily concentrated towards the hot spot, is 
stronger in the northeast, and is enhanced in regions of strong narrow-line 
emission. The stellar component is more extended, concentrated southwest of 
the hot spot, dominated by an old ($\geq$ 2 x 10$^{9}$ years) stellar 
population, and includes a nuclear stellar cluster which is $\sim$200 pc in 
extent.

\end{abstract}


\section{Introduction}

NGC 1068 is the nearest and brightest Seyfert 2 galaxy, and has been the subject 
of intense scrutiny over the entire electromagnetic spectrum (including a recent 
workshop devoted entirely to this object, see  Gallimore \& Tacconi 1997). The 
detection of broad 
(FHWM $\approx$ 4500 km s$^{-1}$) emission lines in the polarized flux from 
NGC 1068 (Antonucci \& Miller 1985; Miller, Goodrich, \& Mathews 1991)
revealed a Seyfert 1 nucleus (characterized by 
broad lines and a strong nonstellar continuum) that is hidden from our direct 
line of sight, but can be seen in reflected light due to electron and/or dust 
scattering in the circumnuclear regions. Thus, NGC 1068 formed the basis of 
unified models for Seyfert galaxies, which postulate that Seyfert 1 and 2 
galaxies are basically the same type of object, but viewed from different angles 
relative to the source of obscuration (e.g., an optically thick torus, see 
Antonucci 1993).

Ground-based images of the outer narrow-line region (NLR), obtained in narrow 
wavebands centered on strong emission lines (Pogge 1988), have revealed a 
cone-like structure along the radio axis (which can also be explained by a thick 
torus).  Subsequent observations with the {\it Hubble 
Space Telescope} (HST) at high spatial resolution ($\sim$0\arcsecpoint1) have 
have shown that the inner NLR contains numerous knots and filaments, 
also in a cone-like geometry (Evans et al. 1991; Macchetto  et al. 1994) with an 
apparent opening angle of 65\deg\ $\pm$ 25\deg\ and a position angle on the sky 
of PA $\approx$ 15\deg\ (whereas the outer NLR has PA $\approx$ 30 -- 
35\deg). These authors admit that the apex of the cone is difficult to determine 
and seems to change with contrast level, but lies within $\sim$0\arcsecpoint4 of 
the continuum peak.

{\it HST} images of the optical continuum emission from the NLR show a bright, 
resolved  peak of emission, which has been named 
the ``continuum hot spot''. The hot spot is $\sim$0\arcsecpoint3 in size and is 
surrounded by an extended structure 
that is elongated in the NE-SW direction, has approximate dimensions of 
3\arcsecpoint5 x 1\arcsecpoint7, and is centered about 0\arcsecpoint4 SW of the 
hot spot. Lynds et al. (1991) suggest that this extended region is dominated by 
stellar light, but that polimetry and/or spectroscopy are needed for 
verification. By contrast, HST images of the UV continuum emission resemble 
those of the emission lines, and in particular, show that the hot spot and a 
conical region NE of it are bright in the UV (Kriss et al. 1993; Macchetto et 
al. 1994).

Numerous spectra of NGC 1068 have been taken from the ground (e.g., see Koski 
1978), and a high-quality UV spectrum obtained with {\it IUE} is shown 
in Snijders, Netzer, \& Boksenberg (1986). Caganoff et al. (1991) give the first 
{\it HST} spectrum, which was obtained by the Faint Object Spectrograph (FOS) 
through a 0\arcsecpoint3 diameter aperture centered on the continuum hot spot, 
and covers the wavelength region 2200 -- 7000 \AA. These authors suggest that 
the hot spot is {\it not} the source of reflected light from the hidden 
nucleus, because they were unable to detect broad H$\beta$ at the large velocity 
width and equivalent width seen in optical spectropolarimetry. 
However, Antonucci, Hurt, \& Miller (1994) present {\it HST} UV (1575 -- 3300 
\AA) spectropolarimetry of the hot spot (see also Code et al. 1993), and show 
that the highly polarized 
($\sim$20\%) light within the 0\arcsecpoint3 aperture shows broad 
C III] and Mg~II emission, as well as blended Fe II emission,
indicating that at least some of the light from the hot spot is reflected 
emission from the hidden broad-line region (BLR).
Antonucci et al. (1994) also find that the scattering region is extended over 
at least $\sim$1$''$, and is dominated by electron scattering on this spatial 
scale, since the polarization does not change as a function of wavelength.

Other FOS observations have been obtained at various locations in the NLR, and 
we have compared these observations with detailed photoionization models to 
study the physical conditions in these regions (Kraemer, Ruiz, \& Crenshaw 
1998). However, these observations were obtained prior to the installation of 
COSTAR, and the regions sampled contained large contributions from outside
the 0\arcsecpoint3 aperture used. 
Clearly, observations at spatial resolutions comparable to that of 
the HST images ($\sim$ 0\arcsecpoint1) are important for understanding the 
nature of the hot spot, as well as the detailed structure and 
kinematics of the NLR in NGC 1068. The Space Telescope Imaging Spectrograph 
(STIS) is an ideal instrument for this purpose, since it provides long-slit 
capability at high spatial resolution in the UV, optical, and near-IR.

In this paper, we describe our STIS observations of NGC 1068, and investigate 
the nature of the continuum emission, including the hot spot and its 
surroundings. In subsequent papers, we give results 
on the compact high-ionization region, the detailed kinematics of the 
emission-line clouds, and the physical conditions and reddening across the NLR.
We adopt a systemic redshift of cz $=$ 1148 km~s$^{-1}$ from H~I observations 
(Brinks et al. 1997) and a distance of 14.4 Mpc (Bland-Hawthorne 1997), so that 
0\arcsecpoint1 corresponds to 7.2 pc.

\section{Observations and Data Reduction}

We observed NGC 1068 with {\it HST}/STIS on 1998 August 15. We used the 52$''$ x 
0\arcsecpoint1 slit to obtain low-dispersion spectra over the wavelength range 
1150 -- 10,270 \AA\ at a spectral resolving power of $\lambda$/$\Delta\lambda$ 
$\approx$ 1000 and spatial resolution of 0\arcsecpoint1 (CCD detector) or 
0\arcsecpoint05 (MAMA detectors, which only contain 25$''$ of the slit length). 
Table 1 gives the grating, detector, wavelength coverage, and total exposure 
time for each setting.

Our slit position was chosen to intersect a number of bright emission-line knots 
in the inner NLR. Figure 1 shows the slit position, at PA $=$ 202\deg, 
superimposed on an FOC [O III] image (Macchetto et al. 1994).
Our procedure for positioning the slit was to perform an 
extended-source target acquision with STIS on the hot spot (using the continuum 
centroid in a 0\arcsecpoint35 x 0\arcsecpoint35 box), and offset the center of 
the slit to a position 0\arcsecpoint14 north of the continuum peak. This 
position places emission knot C in the center of the slit, and includes knots 
B and D (using the nomenclature of Evans et al. 1991); for the purpose of 
discussion, we identify two additional knots in Figure 1 as ``H'' and ``I''.
Figure 2 shows our slit position superimposed on a WFPC2 continuum image 
centered at 5470 \AA. From this image, one can see that the slit includes a 
significant portion of the hot spot, offset from its center.

We reduced the STIS data using the IDL software developed at NASA's Goddard 
Space Flight Center for the Instrument Definition Team. We 
identifed and removed cosmic ray hits using the multiple images obtained in each 
mode, and removed hot or warm pixels (identified in STIS dark images) by 
interpolating over them in the dispersion direction. We used 
wavelength calibration exposures obtained after each science observation to 
correct the wavelength scale for zero-point shifts. In order to correct for the 
high instrumental background of the near-UV MAMA (Kimble et al. 1998), we 
determined a background spectrum (in counts/sec) by median filtering and 
averaging 100 rows near the top of the G230L spectral image (at a projected 
location 9\arcsecpoint5 -- 12\arcsecpoint0 southwest of the hot spot), and 
subtracted this background from each row in the image. We performed the same 
procedure for the other spectral images, although this had little effect, since 
the background counts for these detectors were much lower. Finally, we used the 
``extended'' option in our reductions, which geometrically rectifies the 
spectral images to produce a constant wavelength along each column (in the 
spatial direction), and fluxes at each position along the slit in units of ergs 
s$^{-1}$ cm$^{-2}$ \AA$^{-1}$ per cross-dispersion pixel (fluxes are conserved 
in the geometric transformations).

We show the flux-calibrated images for the G140L and G430L observations in 
Figures 3 and 4; note that the top of these images corresponds to the lower 
right-hand (SW) area in Figures 1 and 2. The images show both smooth and 
discrete structure in the spatial distribution of continuum and emission-line 
fluxes, which appear to coincide in many locations.  (The emission lines show 
complex kinematic structure, which will be the topic of another paper.) The 
continuum hot spot is very prominent in both images, and we will use it as a 
reference location.  The prominent continuum streak below the hot spot is knot 
``H'', and the one above is ``I''. One interesting aspect of these images is 
that the extended UV continuum is concentrated below the hot spot (NE) in Figure 
3, whereas the optical continuum is concentrated above the hot spot (SW) in 
Figure 4.

To obtain spectra at different spatial locations, we used the extraction bins 
shown in Figures 1 and 2. We found that a bin length of 0\arcsecpoint2 (4 CCD 
pixels, 8 MAMA pixels) yields reasonable signal-to-noise ratios in the 
continuum, and is sufficient to isolate the extended continuum and emission-line 
structure. We extracted spectra from each bin to a distance of $\pm$6$''$ from 
the hot spot, and, at each position, combined spectra from different gratings in 
their regions of overlap. The central 0\arcsecpoint2 x 0\arcsecpoint1 bin 
contains the portion of the hot spot that was intercepted by the slit.

\section{Results}

\subsection{Scattered Broad-Line Emission from the Hot Spot}

In Figure 5, we compare the entire UV to near-IR spectrum of the hot spot with a 
STIS spectrum of the nucleus of NGC 4151, obtained over the same wavelength 
region (from Nelson et al. 1999, reduced in 
flux by a factor of 0.015). The overall 
continuum shapes are remarkably similar, and both spectra show a ``little blue 
bump'' in the 2000 - 4000 \AA\ region. The blue end (2200 \AA\ -- 3200 \AA) of 
this feature was detected in NGC 1068 by {\it IUE} (Snijders et al. 1986), and 
confirmed with FOS spectra (Antonucci et al. 1994).
These authors attribute the feature to reflected Fe II emission from the hidden 
nucleus, which is expected from the presence of optical Fe~II emission in 
spectropolarimetry of NGC 1068 (Antonucci \& Miller 1985; Miller et al. 1991).
Our STIS spectra show that this feature extends to 4000 \AA\ and is 
the same shape as that in Seyfert 1 galaxies, which confirms its identity as the 
little 
blue bump. The presence of significant emission above the continuum at 3200 -- 
4000 \AA\ indicates that much of this feature is due to Balmer continuum 
emission, since Fe II contributes little to the bump in this wavelength region 
(Wills, Netzer, \& Wills 1985). The little blue bump is a common feature in
the spectra of Seyfert 1 galaxies and QSOs, and can be explained by a 
combination of Fe II and Balmer continuum emission from dense gas in the BLR 
(Wills et al. 1985). As can be seen in Figure 5, the strength of the little blue 
bump in the hot spot spectrum, relative to the continuum, is 1/3 that in the NGC 
4151 spectrum. Nevertheless, the presence of this feature in our spectrum 
clearly identifies the continuum hot spot as a strong reflector of light from 
the hidden nucleus.

Further evidence for reflected light can be seen in Figure 6, which shows an 
expanded view of the far-UV spectrum of the hot spot. This spectrum shows narrow 
emission lines from knot B (which overlaps with the hot spot) and strong 
absorption lines from our Galaxy, but no evidence for broad stellar absorption 
features like those seen in the N~V and Si~IV lines of the Seyfert 2 galaxy Mrk 
477, which indicate the presence of hot young stars in this object (Heckman et 
al. 1997).
However, broad components are clearly seen in the 
L$\alpha$ and C~IV $\lambda$ 1550 emission lines (which were first detected in 
the {\it IUE} spectrum of Snijders et al. 1986), indicating reflected radiation 
from the hidden nucleus. The broad component 
of C~IV is the easiest to isolate, and is characterized by a velocity width 
(FWHM) $=$ 5200 $\pm$800 km s$^{-1}$ and an equivalent width (EW) $=$ 23 $\pm$2 
\AA. The velocity width is in the middle of the range for Seyfert 1 galaxies, 
whereas the EW is definitely on the low side, but not outside of the Seyfert 1 
range (see Kinney et al. 1990).

Our detection of the little blue bump and broad components of the far-UV 
emission lines confirms Antonucci et al.'s (1994) finding, from near-UV (1600 
-- 3300 \AA) spectropolarimetry, that the hot spot reflects emission from the 
hidden BLR. We can also compare our optical spectra with those obtained through 
a 0\arcsecpoint3 diameter aperture with the FOS (Caganoff et al. 1991), and with 
ground-based spectropolarimetry through a 2$''$-wide slit (Miller et al. 1991). 
Caganoff et al. (1991) reported that they saw a component of H$\beta$ that has a 
FWHM of only 2200 km s$^{-1}$, and concluded that they did {\it not} detect the 
broad component of H$\beta$ seen in the polarized spectra of the nucleus, which 
has a FWHM of 4480 km s$^{-1}$ (Miller et al. 1991). Figure 7 shows our removal 
of the narrow component of H$\beta$ in the STIS spectrum, using the 
[O~III]~$\lambda$5007 line as a template (note that the velocity width of
[O III] is large: FWHM $=$1180 km s$^{-1}$). We find that the velocity width of 
the broad component of H$\beta$ is FWHM $=$ 3800 $\pm$ 700 km s$^{-1}$, which is 
consistent with Miller et al.'s results, to within the errors.

However, we confirm Caganoff et al.'s measurement of a small equivalent width 
for broad H$\beta$: our measurements yield EW (H$\beta$) $=$ 25 
$\pm$5 \AA, compared to Caganoff et al.'s 16 \AA, whereas Miller et al.'s 
spectropolarimetry yields 110 \AA. As we mentioned earlier, the equivalent 
widths of C~IV and the little blue bump are also low, compared to most Seyfert 1 
galaxies, indicating intrinsically weak emission from the BLR.
The small equivalent widths in our spectra cannot be 
be to a large stellar contribution to the continuum, because the hot spot's 
continuum emission is dominated by scattered light; since the stellar light 
contribution  is $\leq$ 31\% (see below), the equivalent width of H$\beta$, 
relative to the nonstellar continuum, cannot be larger than 36 \AA.

Assuming that the starlight has been properly accounted for in the ground-based 
observations, our results indicate that the equivalent width of broad H$\beta$, 
relative to the nonstellar continuum, is a function of aperture size. Since the 
HST spectra are sampling a scattering region that is much smaller 
(0\arcsecpoint3 or 21.6 pc) and closer (on average) to the continuum source than 
that sampled by ground based-observations, a likely explanation is that the 
equivalent width of broad H$\beta$ was actually much lower over a period of 
$\sim$70 years (the light travel time across the hot spot), compared to the 
previous $\sim$700 years. For example, the BLR clouds could be 
matter-bounded, so that a large increase in luminosity would not result in a 
significant increase in broad-line emission. Another 
possibility is that the amount of material in the BLR was much smaller over this 
time period, resulting in a low covering factor, and hence small equivalent 
widths.

\subsection{The Continuum Emission from the Hot Spot}

As we mentioned previously, the far-UV to near-IR continuum of the hot spot in 
NGC 1068 is very similar to that of Seyfert 1 galaxies. To quantify the 
similarity, we fit the continuum regions over 1200 -- 10,000 
\AA\ (excluding the little blue bump) with a power law of the form F$_{\nu}$ 
$\propto$ $\nu$$^{\alpha}$. For the hot spot, we find $\alpha$ $=$ $-$0.5 
$\pm$0.2, whereas for NGC 4151, we find $\alpha$ $=$ $-$0.6 $\pm$0.1. 
The shape of the hot spot continuum is essentially identical to that of 
NGC 4151, and similar to that seen in Seyfert 1 galaxies and QSOs (Wills et al. 
1985). From this 
remarkable similarity, we conclude: 1) the continuum emission from the hot spot 
is dominated by reflected light from the hidden nucleus, since a large stellar 
contribution would produce a much redder continuum (see below), and 2) the 
scattering is wavelength independent and must therefore be due to electrons, 
since dust scattering would produce a much bluer spectrum than generally 
observed in Seyfert 1 galaxies.

There appears to be a slight turnover in the spectrum of the hot spot at 
$\lambda$ $<$ 1450 \AA. Dereddening the spectrum, assuming E(B-V) $=$ 0.1 and 
the standard Galactic reddening curve (Savage \& Mathis 1979), eliminates the 
turnover, but it 1) introduces an artificial emission bump at 2200 \AA, and 2) 
makes the continuum much harder ($\alpha$ $\approx$ $-$0.2) than typical values 
for Seyfert 1 galaxies (Wills et al. 1985). In order to avoid an artificial bump 
at 2200 \AA, the continuum reddening must be E(B-V)~$\leq$~0.05, but this upper 
limit can be avoided if the bump is not a feature of the reddening curve in this 
object (Kriss et al. 1992; Antonucci et al. 1994). Another indication of low 
continuum reddening was found by Kriss et al. (1992) from a power-law plus 
reddening curve fit to the observed continuum in {\it HUT} spectra obtained 
through a large aperture (18$''$), resulting in E(B-V) $=$ 0.065 $\pm$ 0.02.

\subsection{The Extended Continuum Emission}

Figure 8 shows our extracted spectra at four locations along the slit: the hot 
spot, positions H and I, and a region 3$''$ to 6$''$ SW of the hot spot (the 
``galaxy'' spectrum).
It is clear that the continuum shape varies greatly as a function of position, 
suggesting the presence of more than one contributor. In order to separate the 
contributions, we assumed that the continuum at each location consists of two 
components: 1) scattered-light from the hidden nucleus, which is given by the 
spectrum of the hot spot, and 2) stellar light from the host galaxy, which is 
obtained by summing the region 3$''$ to 6$''$ SW of the hot spot. We then 
derived templates for the scattered and stellar continua by fitting these two 
spectra  with cubic splines in regions unaffected by emission or absorption (but 
including the little blue bump in the scattered spectrum). These fits are shown 
in the top and bottom plots of Figure 8. We ignored the ends of the G230L 
spectrum and the blue end of the G430L spectrum in these fits, since they tend 
to be extremely noisy, due to low sensitivity and a high background contribution 
in these regions, as can be seen in the bottom plot of Figure 8.

In order to determine the contribution of the scattered and stellar components 
to the continuum at each location, we took advantage of the fact that the flux 
of the stellar template is essentially zero in the 1400 -- 3000 \AA\ range.
Thus, the contribution of the scattered component at each location is given by 
the UV flux in this wavelength region. After subtracting the scattered light 
component, scaled to match the observed UV flux, the contribution of the stellar 
component is then given by the residual flux in the optical. The success of this 
simple procedure is demonstrated by our fits to positions H and I in Figure 8, 
and, in general, the two components provided a good match to the observed 
continuum at all locations. The discrepancies 
between observed and model continuum shapes are small, and may be the result of 
slight variations in the reddening along the slit. We conclude that the extended 
continuum is well represented by a combination of scattered and stellar 
components, whose relative contributions vary as a function of position.

Using the above procedure, we derived brightness profiles along the slit for the 
scattered and stellar components at 5500~\AA; these profiles are plotted in 
Figure 9. The scattered light is heavily concentrated towards the hot spot,
but also shows an extended component over the $-$6$''$ (NE) to $+$2$''$ (SW) 
range. The scattered light is much stronger in the region NE of the hot spot, 
including a large bump from $-$2\arcsecpoint2 to $-$0\arcsecpoint4. There are 
also discrete sources of scattered light in addition to the hot spot; the most 
prominent of these are at positions H and I. These features can also be seen in 
our far-UV spectrum (Figure 3), since the UV is dominated by scattered light.

In Figure 10, we plot the brightness profiles of the scattering continuum and 
the [O~III] flux along the slit, normalized to their maximum values. The 
discrete structures in these profiles, represented by the peaks, coincide, 
although the amplitudes differ. In particular, emission-line knots B (in the red 
wing of C), H, and I correspond to peaks in the scattered light profile, and 
knots C and D correspond to the bump in the wing of the scattered light profile. 
This indicates that regions of ionized gas responsible for the emission lines 
are approximately co-located with regions responsible for the enhanced electron 
scattering, although they are not necessarily identical.

The brightness profile of the stellar component in Figure 9 is smoother and less 
concentrated than that of the scattered component. In addition to a uniform 
stellar flux over this region, there is a large bump within $\pm$2$''$ the hot 
spot, which is stronger to the NW; the derived peak of the stellar flux is at 
$+$0\arcsecpoint2. Since the optical continuum receives a 
large contribution from the stellar flux, these features can also be seen in the 
optical spectrum in Figure 4. 
Our assumption that the spectrum of the hot spot is entirely due to scattered 
light precludes an accurate determination of the stellar contribution at this 
location. However, we note that the centroid of the stellar profile is 
$\sim$0\arcsecpoint3 NW of the hot spot (close to the observed peak), and it is 
therefore likely that the stellar flux at the hot spot is less than the peak. 
At 5500 \AA\, the total continuum flux at the hot spot is 6.5 x 10$^{-16}$ ergs 
s$^{-1}$ cm$^{-2}$ \AA$^{-1}$, and the stellar flux is $<$2.0 x 10$^{-16}$ ergs 
s$^{-1}$ cm$^{-2}$ \AA$^{-1}$, so the contribution of the stellar flux to the 
hot spot is therefore $<$ 31\%. This conclusion is consistent with Antonucci et 
al.'s (1994) finding that that the continuum and broad-line polarizations are 
the same in small apertures (0\arcsecpoint3 -- 4\arcsecpoint3 x 
1\arcsecpoint4) centered on the hot spot, indicating the lack of significant 
dilution of the continuum by starlight.

Figure 11 shows expanded views of the host galaxy spectrum from the bottom 
plot in Figure 8. The weakness of the UV flux in this spectrum indicates the 
absence of a significant population of hot young stars. To estimate the age of 
the population, we used the stellar population models and synthetic spectra 
developed by Bruzual \& Charlot (1993). Figure 11 reveals that a reasonably good 
match is obtained with a Salpeter inital mass function (IMF),  a mass range of 
0.1 -- 125 M$_{\odot}$ and an instantaneous starburst with an age of 2 Gyr.
An age $\geq$3 Gyr provides a better match for the flat continuum in the 8000 -- 
10,000 \AA\ region, but a slightly worse fit (lower flux) in the 3000 - 5000 
\AA\ region. Significantly smaller ages (i.e., $\leq$ 1.5 Gyr) produce too much 
flux in the 3000 -- 5000 \AA\ region and much steeper continua in the 7000 - 
10000 \AA\ band. Changes in the synthetic spectra are subtle for greater ages or 
lower mass cutoffs. Given the noisiness of the observed spectra and the 
reasonably good fits in Figure 11, we place a conservative lower limit of 2 Gyr 
on the age of this stellar population.

There is a small amount of UV flux in Figure 11 that cannot be 
explained by this model. This feature does not resemble the smooth ``UV upturn'' 
seen in many elliptical galaxies, which starts around 2000 \AA\ and smoothly 
rises to $\sim$1200 \AA\ (and is thought to be due to hot evolved stars,
such as extreme horizontal branch stars, see O'Connell 1999). Other 
explanations for the small UV excess include a minority 
population of young stars or a small amount of scattered light from the hidden 
nucleus. However, our main conclusion is still valid: the stellar light is 
dominated by a old stellar population with an age $\geq$ 2 Gyr.

\section{Discussion and Summary}

Our detection of the little blue bump and broad emission lines in the spectrum 
of the hot spot proves that this compact region is a strong source of reflected 
radiation from the hidden nucleus, in agreement with the conclusions of 
Antonucci et al. (1994). The velocity widths of the broad components 
are not a problem for this conclusion: our FWHM of broad H$\beta$ is 
substantially larger than the 
measurement of Caganoff et al. (1991), and the same (to within the errors) as 
that reported by Miller et al. (1991) for the nucleus. The FWHM of C~IV is 
larger than that of H$\beta$ by $\sim$1400 km s$^{-1}$, which is a 
common occurence in Seyfert 1 galaxies (Corbin \& Boroson 1996). We confirm 
Caganoff et al.'s measurement of a small equivalent width for H$\beta$, and 
attribute it to an actual variation of equivalent width (relative to the 
nonstellar continuum) with aperture size, indicating a decrease in the 
equivalent width over time. 
The small equivalent widths of the C~IV line and the little blue bump, compared 
to typical Seyfert 1 galaxies, are additional indications of 
intrinsically weak broad emission reflected by the hot spot.

The shape of the hot spot's UV to near-IR continuum is similar to 
that of Seyfert 1 galaxies, which suggests that we are seeing the intrinsic 
shape of the 
continuum, and indicates that 1) the continuum of the hot spot is dominated by 
scattered light, since  a large stellar contribution would significantly redden 
the spectrum, and 2) the reflection that we see is due to electron scattering, 
since dust scattering would ``bluen'' the spectrum. We estimate that the 
scattered light contributes more than 69\% of the hot spot's continuum.

We have decomposed the extended continuum emission in the NLR into two 
components: scattered light and stellar light from the host galaxy. The 
scattered light contributes essentially all of the UV continuum, and is enhanced 
in regions of line emission. This explains why the FOC UV images of Macchetto et 
al. (1994) strongly resemble those in the light of [O III]. It is also 
consistent with Capetti et al.'s (1995) finding that the UV light
must be due to scattering, since it is highly polarized (20 -- 65\%)
over the entire region. In order to 
determine the total contribution of the hot spot to the scattered light, we 
retrieved Macchetto et al.s' (1994) UV continuum image centered on 3520 \AA. 
From this image, we find that the hot spot accounts for 20\% of the UV flux, 
and therefore 20\% of the scattered light in the inner 500~pc.

We have determined that the continuum of the host galaxy in a region 3$''$ -- 
6$''$ SW of the hot spot is dominated by an old ($\geq$ 2 x 10$^{9}$ years) 
stellar population. We have also confirmed the Lynds et al.'s 
(1991) suggestion that the high-contrast structure surrounding the nucleus 
(Figure 2) is stellar. Although the morphology of this structure (Figure 2) is 
somewhat peculiar, its high contrast indicates that it is a stellar cluster.
Our ability to match the continuum in this region with a combination of the 
scattered light and stellar light further out in the galaxy indicates that it is 
an old cluster, with an age that is similar to that derived above.  Additional 
evidence for the lack of young stars in this cluster is the absence of this 
structure in the UV images (Macchetto et al. 1994), and the absence of stellar 
features in our UV spectra. Thatte et al. (1997) have 
detected this nuclear stellar cluster from stellar CO features in the H and K 
bands. Their estimate of its size (FWHM $\approx$ 50 pc) is consistent with the 
width of the stellar light profile in Figure 9 (FWHM $\approx$ 60 pc). Thatte et 
al. estimate the age of the cluster to be 5 -- 16 x 10$^{8}$ years, which is 
slightly less than our estimate for the region further out, but consistent with 
our statements that the cluster and surrounding region are dominated by old 
stellar populations. 

Finally, there is no evidence for another continuum component in our data, which 
agrees with Antonucci et al.'s (1994) conclusion that there is no additional 
``featureless continuum'' in the nuclear region of NGC 1068. Tran (1995) 
concludes that a featureless continuum (other than the scattering component) is 
present in a number of Seyfert 2 galaxies, and Heckman et al. (1995, 1997) 
indicate 
that this feature is probably due to hot stars from recent nuclear starbursts. 
In NGC 1068,  however, the closest starburst  to the nucleus is at a distance of 
$\sim$10$''$ (720 pc), and none of the prominent starbursts (Neff et al. 1994) 
are in our slit. Thus there is no clear connection between current AGN 
activity and recent star formation in NGC 1068.

\acknowledgments
We thank Tom Brown for help on stellar populations and Fred Bruhweiler for 
useful discussions on the STIS spectra. We acknowledge support from NASA grant 
NAG 5-4103.

\clearpage

\figcaption[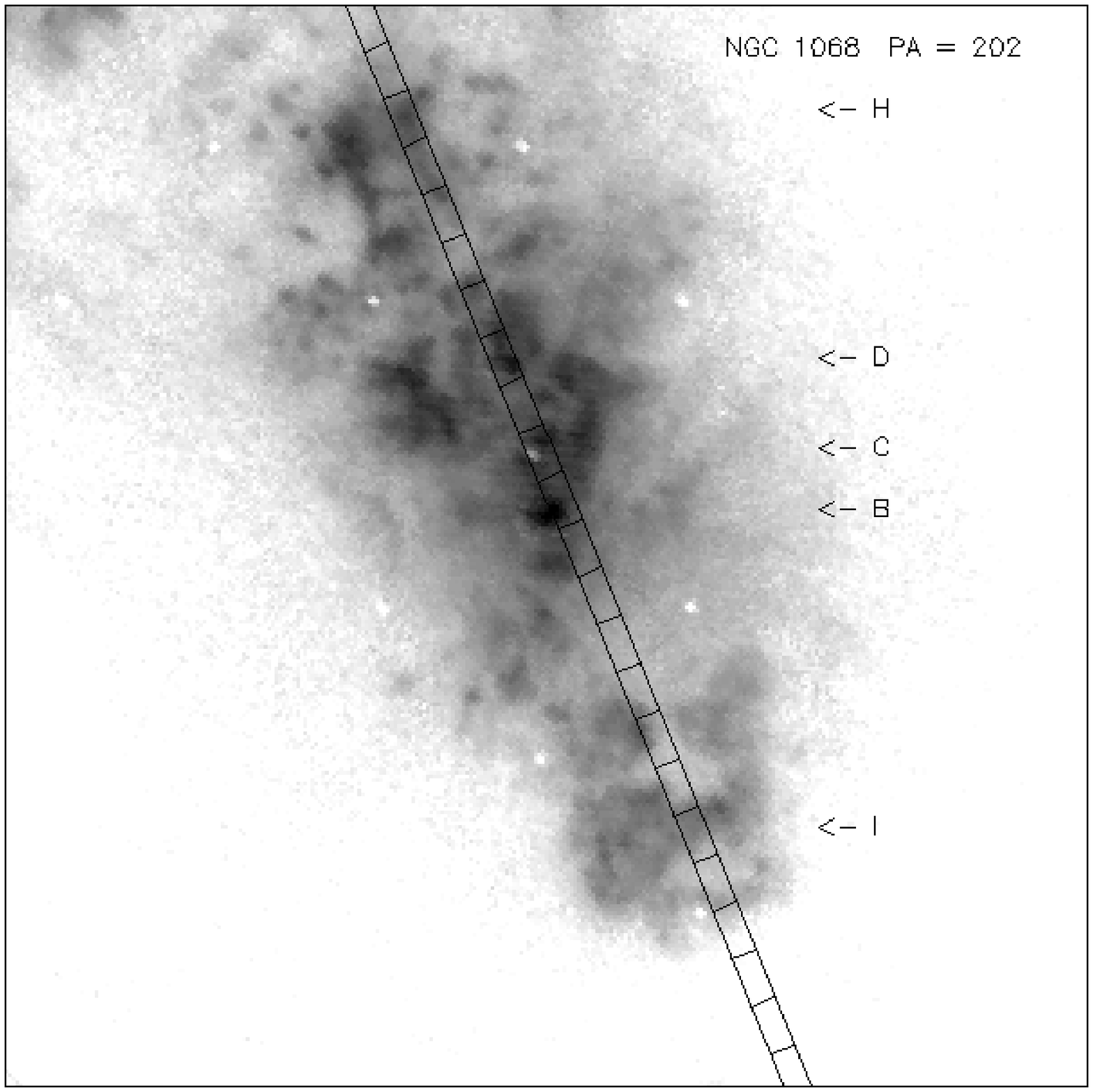]{Slit Position for the low-dispersion spectra in this 
paper, at PA = 202 \deg, superimposed on an FOC [O III] image (the regular 
pattern of white marks are reseaux). Bins (0\arcsecpoint2 x 0\arcsecpoint1) used 
for the extraction of spectra are also plotted. North is up and east is to the 
left.}

\figcaption[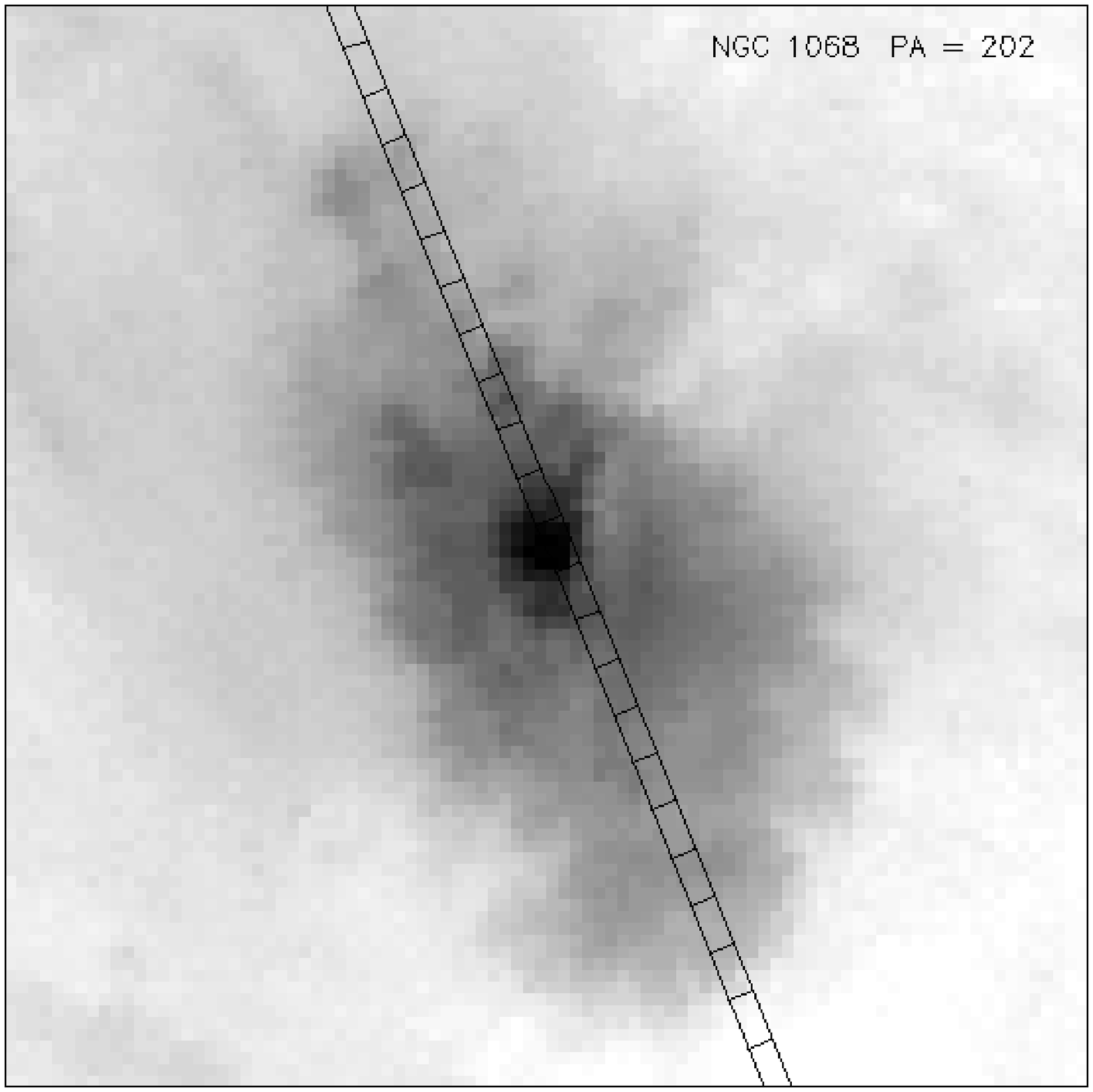]{Slit Position superimposed on a continuum WFPC2 image 
(centered at 5470\AA). The spatial scale and orientation are the same as in 
Figure 1.}

\figcaption[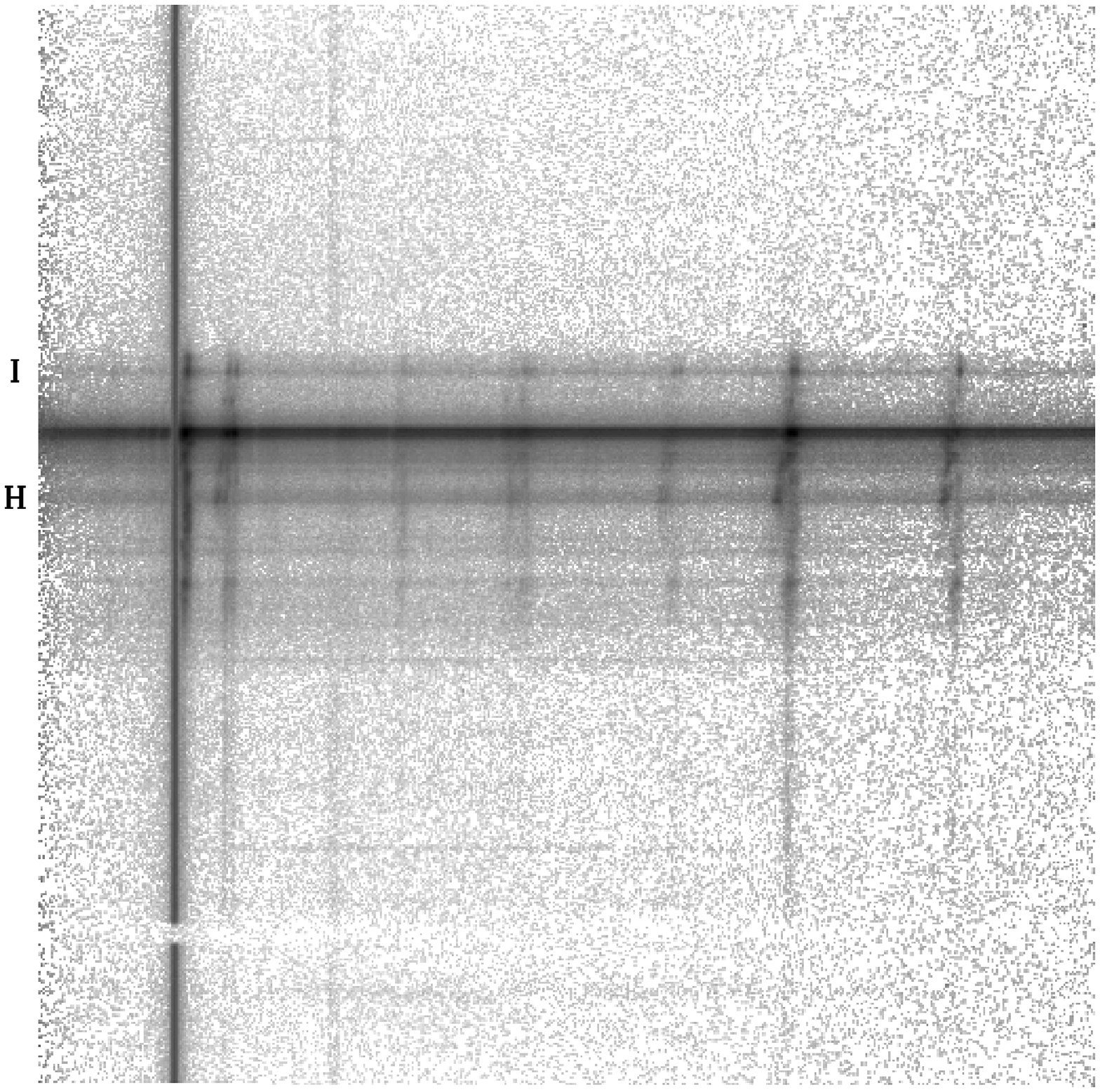]{G140L long-slit image of NGC 1068 at PA $=$ 202\deg; 
wavelength increases to the right and SW is at the top of the image. The 
locations of emission knots H and I are noted.}

\figcaption[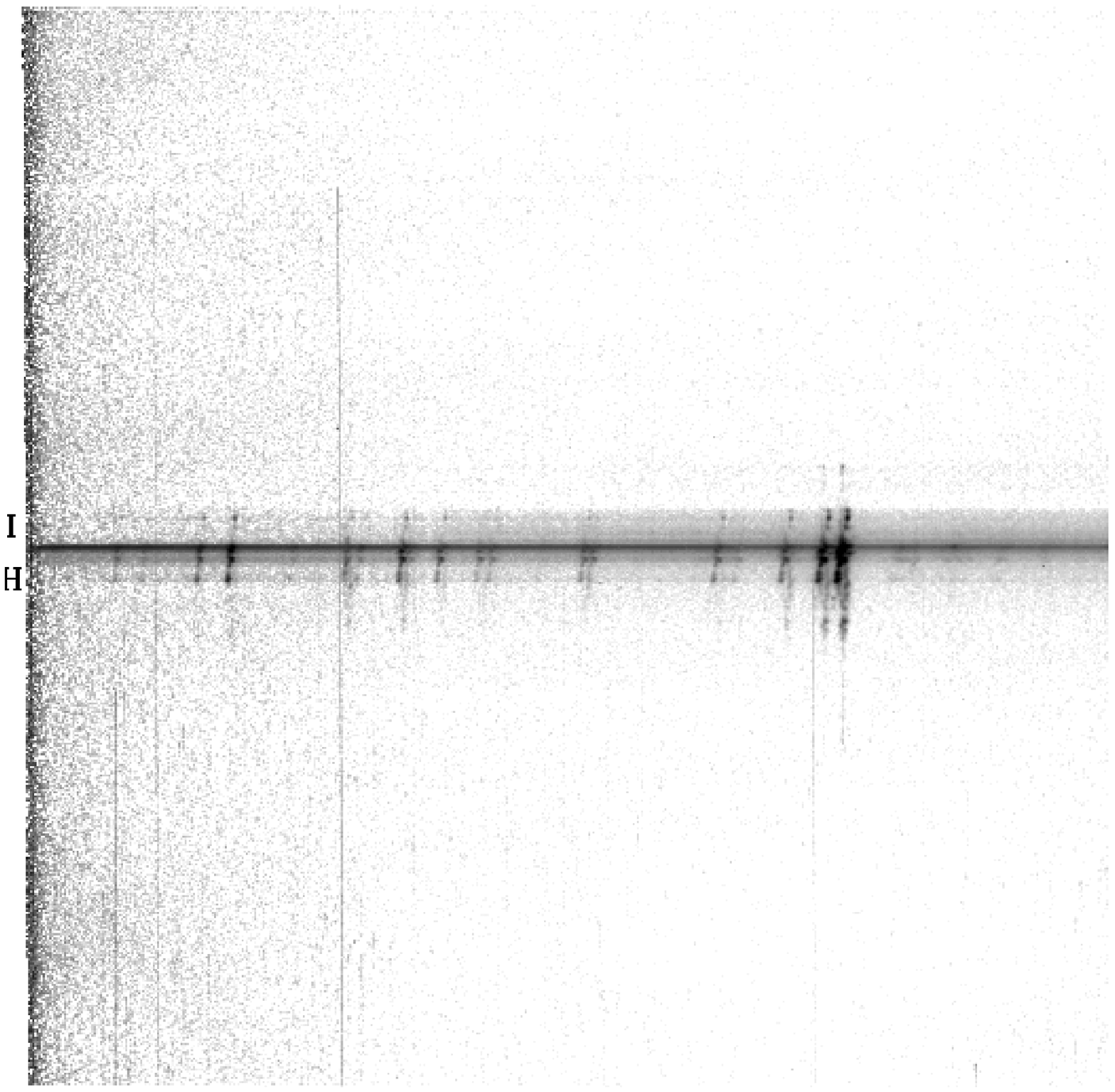]{G430L long-slit image of NGC 1068 at PA $=$ 202\deg;
wavelength increases to the right and SW is at the top of the image.
The spatial scale is compressed by a factor of 1/2 compared to Figure 3, due to 
the larger plate scale for the CCD detector. The locations of emission knots H 
and I are noted. (The vertical lines are bad CCD columns.)}

\figcaption[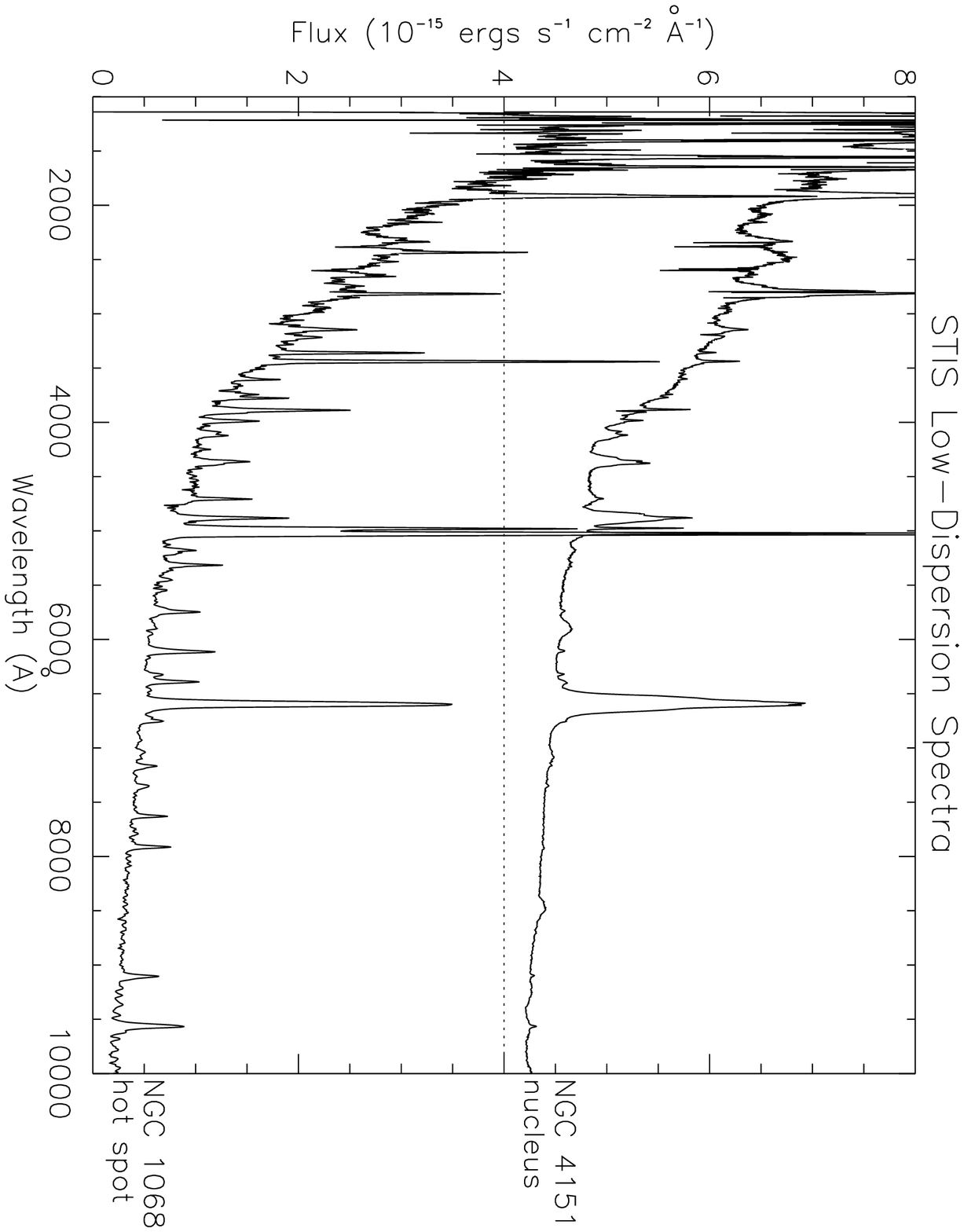]{Comparison of continuum spectrum of the NGC 1068 hot spot 
(smoothed with a 3-point boxcar) with that of the nucleus of NGC 4151 (which has 
been scaled down in flux by a factor of 0.015).}

\figcaption[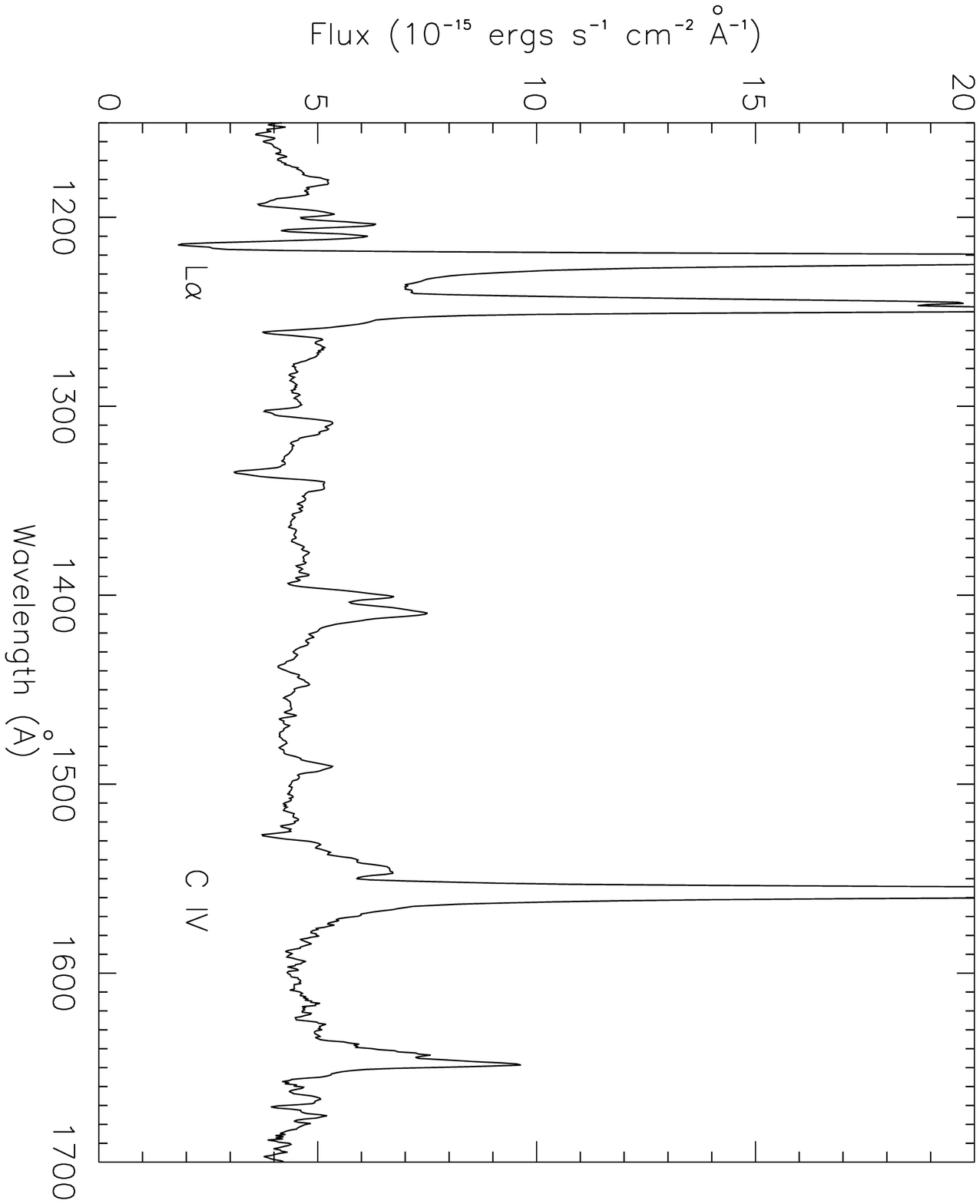]{Far-UV spectrum of the NGC 1068 hot spot (smoothed with a
3-point boxcar). Note the broad components of L$\alpha$ and C IV $\lambda$1550.}

\figcaption[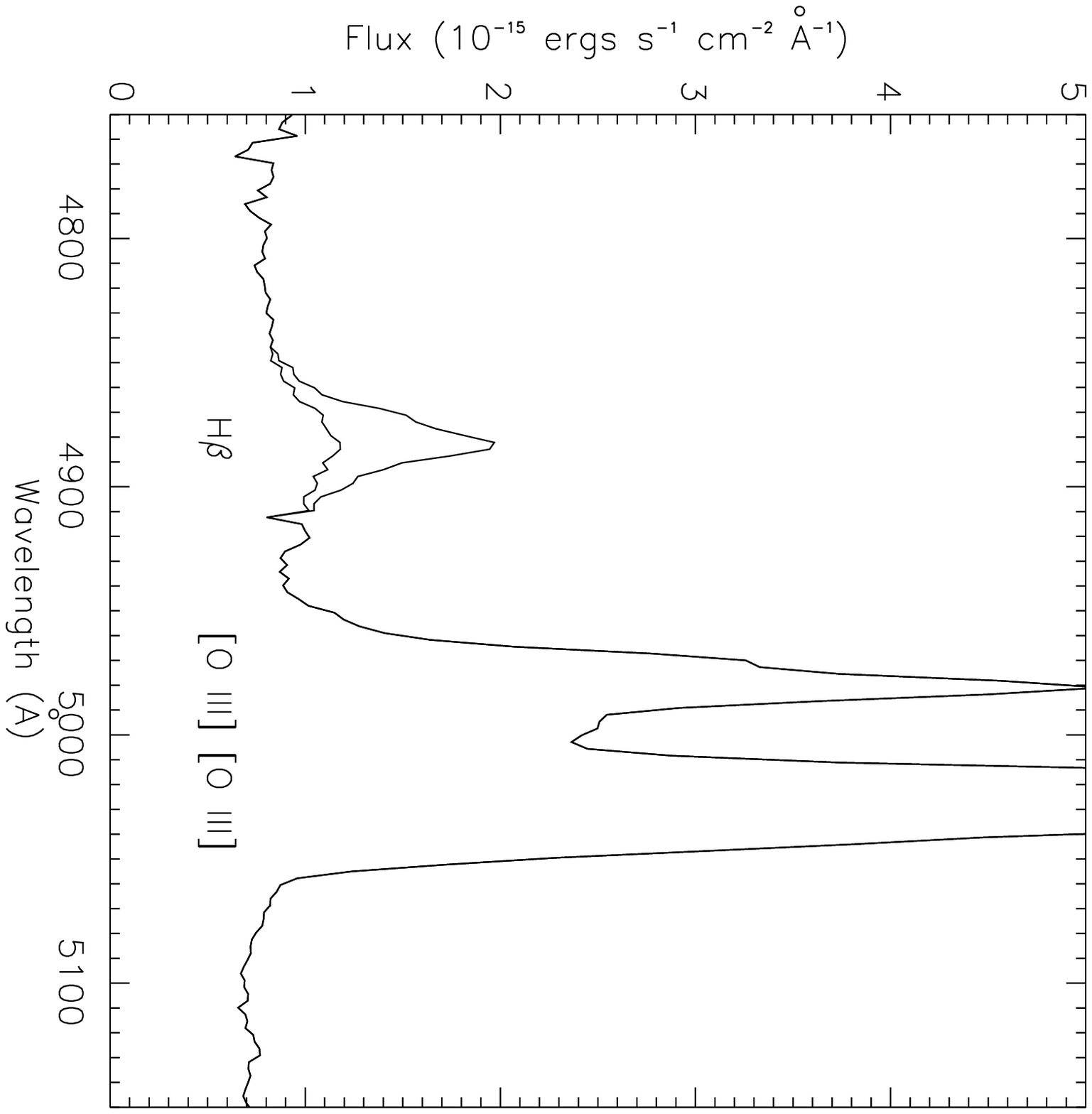]{Spectrum of hot spot in the region of H$\beta$ (upper: 
observed, lower: narrow H$\beta$ removed).}

\figcaption[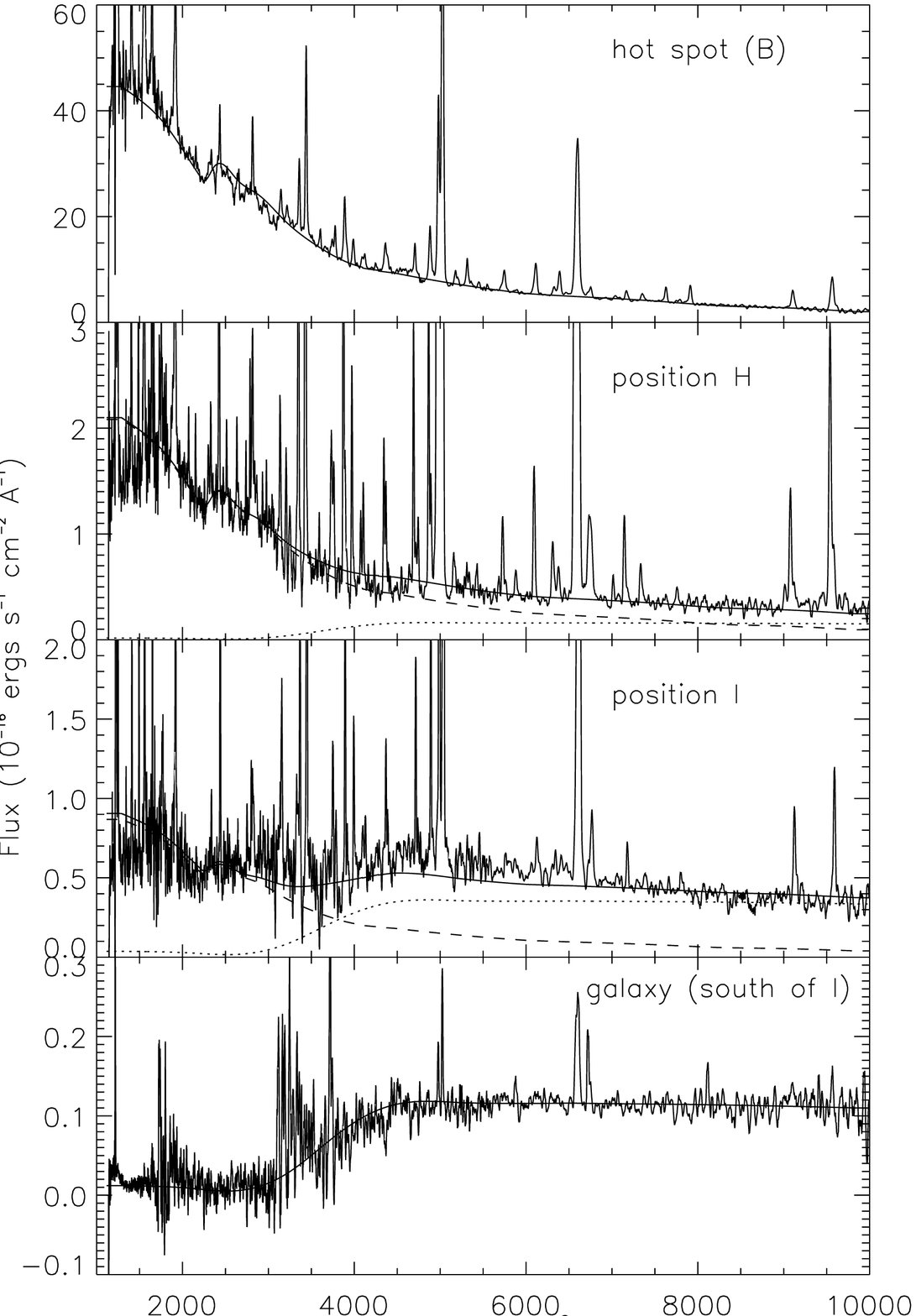]{UV to near-IR spectra at four positions along the slit 
(smoothed with a 5-point boxcar):
the hot spot, location H ($-$1\arcsecpoint5 to $-$1\arcsecpoint7), location I 
($+$1\arcsecpoint3 to $+$1\arcsecpoint5), and the galaxy ($+$3\arcsecpoint0 to 
$+$6\arcsecpoint0).
Continuum fits are shown (solid: total, dashed: scattered component, dotted: 
stellar component).}

\figcaption[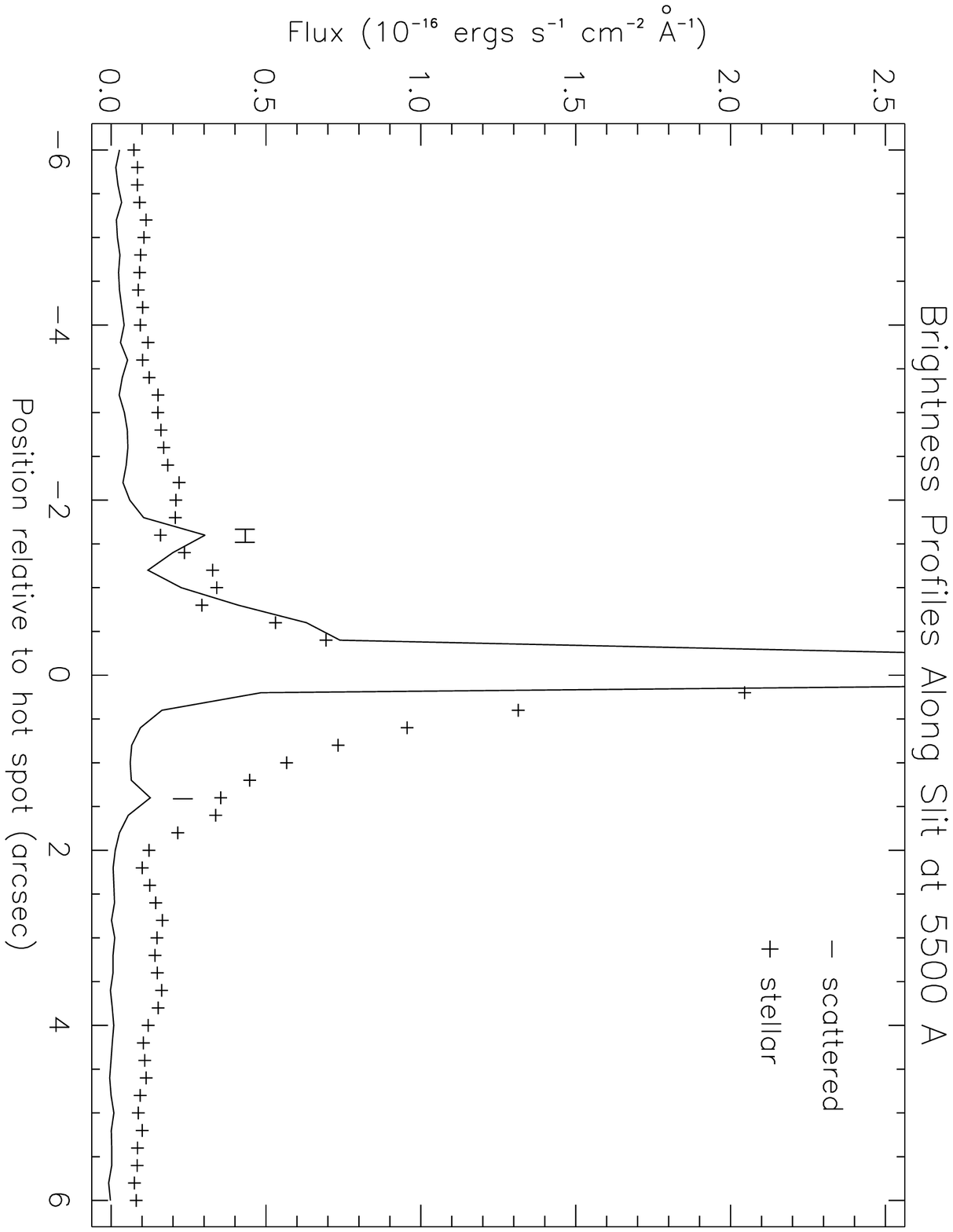]{Brightness profiles of the scattered and stellar continuum 
components along the slit. Negative positions correspond to the NE direction.}

\figcaption[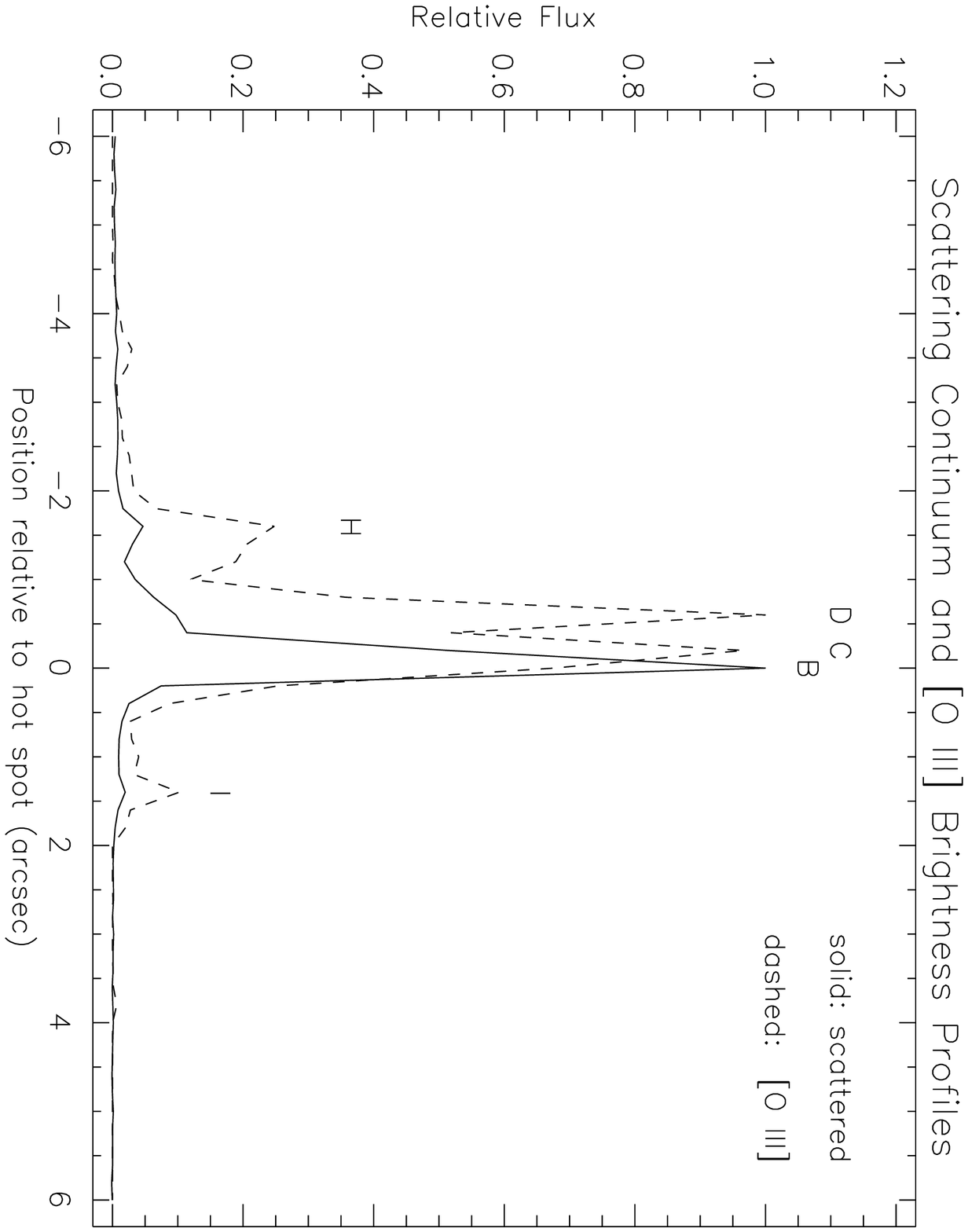]{Relative brightness profiles of the scattered continuum 
and [O III] emission along the slit. Negative positions correspond to the NE 
direction.}

\figcaption[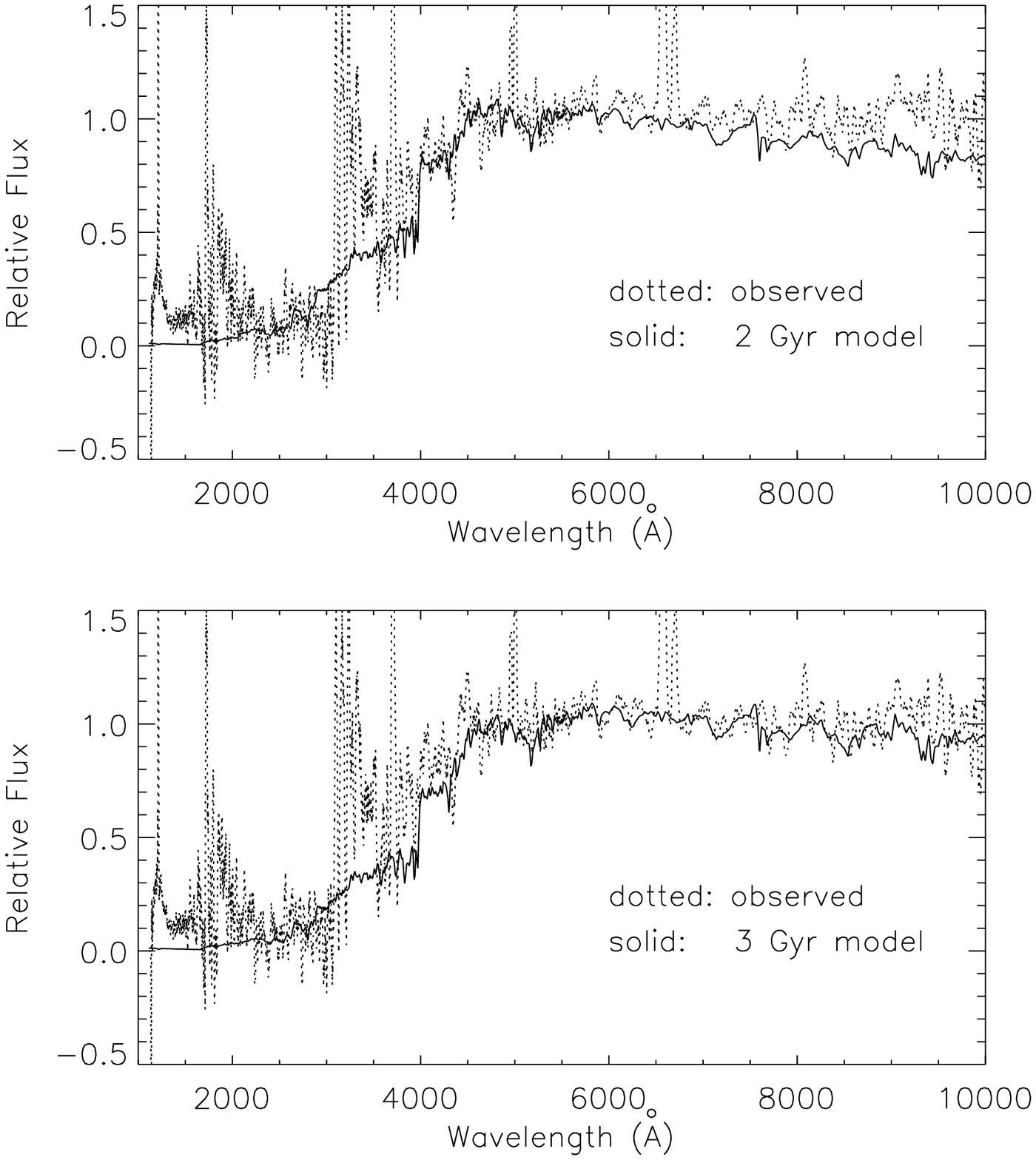]{Comparison of host galaxy spectrum (dotted line, smoothed 
with a 5-point boxcar) with 
synthesized spectra from stellar population models, characterized by 
instantaneous starbursts that occured 2 Gyr (solid line) and 3 Gyr (dashed line) 
ago.}

\clearpage
\begin{deluxetable}{lccr}
\tablecolumns{4}
\footnotesize
\tablecaption{STIS Long-Slit Spectra at PA $=$ 202\deg \label{tbl-1}}
\tablewidth{0pt}
\tablehead{
\colhead{Grating} & \colhead{Detector} &
\colhead{Coverage} & \colhead{Exposure} \\
\colhead{} &\colhead{} & \colhead{(\AA)} & \colhead{(sec)}
}
\startdata
G140L  &far-UV MAMA  &1150 -- 1724   &4736 \\
G230L  &near-UV MAMA &1592 -- 3176   &2280 \\
G430L  &CCD          &2905 -- 5715   &540  \\
G750L  &CCD          &5273~ -- 10,268 &540  \\
\enddata
\end{deluxetable}


\pagestyle{empty}

\clearpage
\begin{figure}
\plotone{fig1.ps}
\\Fig.~1.
\end{figure}

\clearpage
\begin{figure}
\plotone{fig2.ps}
\\Fig.~2.
\end{figure}

\clearpage
\begin{figure}
\plotone{fig3.ps}
\\Fig.~3.
\end{figure}

\clearpage
\begin{figure}
\plotone{fig4.ps}
\\Fig.~4.
\end{figure}

\clearpage
\begin{figure}
\plotone{fig5.eps}
\\Fig.~5.
\end{figure}

\clearpage
\begin{figure}
\plotone{fig6.eps}
\\Fig.~6.
\end{figure}

\clearpage
\begin{figure}
\plotone{fig7.eps}
\\Fig.~7.
\end{figure}

\clearpage
\begin{figure}
\plotone{fig8.eps}
\\Fig.~8.
\end{figure}

\clearpage
\begin{figure}
\plotone{fig9.eps}
\\Fig.~9.
\end{figure}

\clearpage
\begin{figure}
\plotone{fig10.eps}
\\Fig.~10.
\end{figure}

\clearpage
\begin{figure}
\plotone{fig11.eps}
\\Fig.~11.
\end{figure}

\end{document}